\documentclass[preprint2,11pt]{aastex}
\usepackage{epsfig}

\def\beq{\begin{eqnarray}}
\def\eeq{\end{eqnarray}}
\def\upsh{\Upsilon_{\rm HI}}    
\def\NHI{{\rm N_{HI}}}
\def\cm2{{\rm{cm}^{-2}}}  

\def\nth{{\rm N}^{T}_{\rm HI}}
\def\rmd{{\rm d}}
\def\nqso{n_{\rm QSO}}
\def\lya{${\rm Ly-}\alpha~$}
\def\npqso{$n'_{\rm QSO}$}
\newcommand{\nbody}{$N$-body~}

\title{Lensing by Lyman Limit Systems:\\
 Determining the Mass to Gas Ratio}

\author{Ariyeh H. Maller, Tsafrir S. Kolatt}
\affil{Racah Institute for Physics,The Hebrew University, Jerusalem, 
91904 ISRAEL}
\author{Matthias Bartelmann}
\affil{Max-Planck-Institut f\"{u}r Astrophysik,  P.O. Box 1317, 
D-84741 Garching, GERMANY}
\and
\author{George R. Blumenthal}
\affil{UCO/Lick Observatory, Department of Astronomy \&
Astrophysics, University of California, Santa Cruz, CA 95064}

\begin{abstract}
We present a new method to determine $\upsh$, the total
mass-to-neutral gas ratio in Lyman-limits systems.
The method exploits the relation between the neutral hydrogen 
column density and the magnification of background sources due to the
weak gravitational lensing that these systems induce. Because weak
lensing does not provide a direct measure of mass, one must use this
relation in a statistical sense to solve for the 
average $\upsh$ and its distribution.
We use a detailed mock catalog of quasars (sources) and Lyman-limit
systems (lenses) to demonstrate the applicability of this approach 
through our ability to recover the parameter $\upsh$.
This mock catalog also allows us to check for systematics in the 
method and to sketch its limitations.
For a universal constant $\upsh$ and a sample of $\nqso$ quasars,
we obtain an unbiased estimate of its value with $95\%$ confidence 
limits (independent of its  actual value) of
$\upsh \pm 140 \, \sqrt{10^5/n_{\rm QSO}}$.
\end{abstract}


\begin{document}

\section{Introduction}
\label{sec_intro}

The Lyman limit systems  of neutral hydrogen column density 
$\NHI>2\times10^{17} \cm2$ are believed to be closely 
related to galaxies at all redshifts. Among these the rare damped 
\lya systems (DLAS) with $\NHI>2\times10^{20} \cm2$
are usually identified with galactic disks
(\citet{wolfe:86}; see, however, \citet{rhs:97}),
while lower column density systems are more likely to correlate with 
galactic or proto-galactic halos \citep{bs:69,chur:96,mm:96,am:98}. 
Halos are the 
basic building blocks of structure formation when it comes to dark 
matter (DM) gravitational collapse and virialization in any 
hierarchical scenario.
Hence, low column density Lyman limit systems (hereafter simply LLS) 
provide a unique opportunity to study the connection between the DM 
component and the gas component at different redshifts
while the two are still in hydrodynamical equilibrium.

The relation between the DM component and the gas component in LLS 
can be characterized in terms of the ``mass-to-gas'' ratio, in 
analogy to the  mass-to-light ratio commonly used for galactic disks.
In the large scale structure picture this ratio represents
the mean value of the biasing factor, which connects 
the spatial distribution of the baryons
(gas as well as stellar disks) with that of the dark matter.
Generally, the biasing may be a 
linear relation between the gas over-density
and the DM over-density, or a more elaborated functional connection
(e.g., \citealt{dl:99}). 
Biasing can be deterministic, but in reality it is more likely
to be stochastic.
It is possible to calculate such parameters
as the mean biasing in any scheme in which there is a functional 
relation between the bias and the overdensity.

In order to determine the mass-to-gas ratio, one has to observe at 
least two quantities, where one of them is directly related to the 
gas density and the other probes the DM density field. 
This has been in part applied by \citet{croft:98} who presented a 
method by which they recover the DM density fluctuations from the 
\lya forest. A detailed comparison to \nbody\ simulations provided 
them with the ``experimental" bias parameter and spared the need to 
solve for it independently. Based on these simulations,
\citet{kwhm:96} and \citet{hkwm:96} previously realized that 
\lya absorption systems consist of a variety of physical objects: 
collapsing halos, cusps in phase
space for density shells, and galactic disks. The nice agreement 
between the results from the simulations and the observed column 
density function added credibility to this picture.

The appeal to \nbody\ simulations can be computationally
expensive, and it is unclear whether the higher resolution needed 
for the LLS can be achieved today in a large cosmological simulation 
\citep{gard:99}. The simulations also make assumptions on the 
physical conditions and the equation of state of the gas, which also 
limits their use. It is therefore desirable to construct an 
independent method by which the mean mass-to-gas ratio and other 
parameters can be measured.

Here we propose to use gravitational lensing to determine the 
matter content, baryonic fraction and ionization state of 
the LLS.
The absorption features, being a direct probe of the {\it neutral} 
hydrogen content in LLS, provide complementary data needed to reveal 
the connection between the matter and HI densities.

The absorption features are well known and well measured 
\citep[see][for recent reviews]{petit:98,rauch:98}. Their 
interpretation, although subject to uncertainties in fitting the
line profile, is rather straightforward and can be expressed in terms 
of their column density $\NHI$, their equivalent width, and their 
velocity parameter. The lensing due to the moderate column densities 
of the LLS has not previously been considered because of the low 
signal expected.

So far only lensing by DLAS ($\NHI>2\times10^{20} \cm2$) has been 
studied for its effect on the magnification bias for the background 
source population, as well as on the distribution of the observed 
impact parameters with respect to the center of the lens 
\citep{scs:97,bl:96,smet:95}.
Lensing by column densities of $\NHI<10^{20} \cm2$ seemed 
unreasonable to consider because of the very small effect it has 
on the background sources; a typical quasar is only magnified by
about one percent. Theoretically, only the overall effects of lensing 
on quantities like the luminosity function of quasars have been 
considered \citep{hama:00,pei:95b}, while observationally a possible 
excess in the number counts of metal line systems has been detected 
\citep{vqy:96}. 

Other manifestations of weak lensing have been detected and used to 
map the mass distribution in clusters and larger scale structure
(see \citealt{bs:00} for a review). 
The standard lore in weak gravitational lensing is to use the 
ellipticity as induced by the intervening gravitational shear field 
in order to resolve the projected mass distribution along the line 
of sight \citep{ks:93}. 
Due to the small level of magnification and the unavailable absolute 
luminosity of the sources, the observed flux of the background sources
has never been used in weak lensing studies. For this reason, point 
sources, for which ellipticities cannot be measured, cannot be used 
to map intervening mass concentrations in this fashion.
The weak magnification of point sources has usually been considered 
only as a nuisance in extracting information from the background 
sources (e.g., in the case of supernovae Ia \citealp{wamb:97,pm:00}).

Until now, a key aspect of the phenomenon of weak lensing by LLS has 
been neglected, the fact that the {\it same} systems that cause 
the magnification are also responsible for the troughs in the 
continuum emission of the background source. Both are proportional 
to projected surface densities along the line of sight. The two are 
therefore correlated. It is this correlation, and the fact that it 
holds for each individual system that we wish to 
exploit in order to extract statistically the relation between the
total mass of these systems and their HI content. In addition,
the absorption system gives the redshift of the lens.

In a large sample, the magnification of point sources can have a
strong statistical effect. Treating all quasars as a set eliminates 
in part the need to know the absolute magnitudes of the individual 
sources. The magnification shifts the luminosity function of the 
combined population in a calculable way. The lensed population can be
``de-magnified'' (assuming a mass-to-gas scheme) on an individual 
basis. The criterion for a successful correction is the return to the
statistical properties of an un-lensed population.

Every statistical approach requires a large 
data base in order to maximize the signal to noise ratio.
Fortunately in the upcoming years such a data base will become
available when the quasar sample from the Sloan Digital Sky Survey 
(SDSS) is completed. This sample should serve as an excellent 
target list by itself or for follow-up observations.
Combined with high resolution spectra for a selective sub 
sample from ground- and space-based telescopes it will allow 
coverage of \lya absorption features over a wide range of redshifts.

Here, we propose a recipe for an observational project 
to determine the mass-to-neutral-hydrogen ratio in LLS, 
check different bias schemes, or apply a similar technique to the 
mass-to-metal-line ratio in these systems.

In \S \ref{sec_theory} we define our notation and present the 
formalism of the technique. In order to describe the main features 
of the proposed method in \S \ref{sec_sm}, we introduce a toy model 
which avoids many complications and which allows us to outline
the ingredients of the technique. 
The next section (\ref{sec_method}) is devoted to the generation of 
a realistic mock quasar and Lyman-limit system catalog which enables
various estimates of the method's biases and errors, as well as the
functional dependence of the obtained signal-to-noise ratio on
the input parameters. In the same section, we use Monte-Carlo
realizations of that catalog and describe our procedure for 
measuring the mass-to-gas ratio. The method 
is implemented, and its advantages and limitations are discussed. 
We conclude this section by additional checks that verify that the 
measured effect indeed comes from gravitational lensing. In section 
\ref{sec_observe} we sketch the preferred procedure
for actual observations, consider extensions of the method, and 
estimate additional effects like dust.  Section \ref{sec_conc} 
summarizes the method's strengths and weaknesses. 

\section{The Theory} \label{sec_theory}
It is common in gravitational lensing to express the magnification 
in terms of two dimensionless quantities; $\kappa$, the convergence 
and $\gamma$, the shear.  The magnification $\mu$ is then given by
\beq
\mu={{1}\over{(1-\kappa)^2-\gamma^2}} \,,
\eeq
\citep{nb:99,sef}.
In the weak lensing limit, where $\kappa \sim \gamma \ll 1$ this
equation becomes 
\beq
\mu=1+2\kappa\,.
\eeq
The convergence, $\kappa$, is the projected surface density, 
$\Sigma$, along the line of sight divided by the critical surface 
density,
\beq \label{eq_sc}
\Sigma_{\rm cr}={{c^2}\over{4\pi G}} D_{\rm eff}\,;\,
D_{\rm eff}={{D_l D_{ls}}\over{D_s}},
\eeq
where $D_{l,s,ls}$ are the angular diameter distances between 
observer, lens and source. 

For absorption systems the column density in neutral hydrogen, 
$\NHI$, along the line of sight is related to $\Sigma$ via 
\beq
\NHI={\rm Y}_{\rm HI} {f_{\rm H} \over m_{\rm H}} f_b f_g \Sigma \, , 
\label{eq_NHI}
\eeq
where $f_{\rm H}$ the fraction of
baryons that are in hydrogen, $m_{\rm H}$ is the hydrogen mass,
$f_b$ is the fraction of total mass
made up by baryons, and $f_g$ is a geometry factor that relates
the possible different geometries of the baryons and the dark
matter (e.g., gas disk in a spherical dark matter halo, etc.).
The fraction of neutral hydrogen is defined by
\beq
{\rm Y}_{\rm HI} = {\rho_{\rm HI} \over \rho_{\rm H}} = 
{\NHI \over {\rm N}_{\rm H}} \, ,
\eeq
where in the last equality we explicitly assume that the volume density
fraction is identical to the surface density fraction.  The geometry 
factor, $f_g$ is of order unity if LLS can be adequately described 
by a roughly spherical halo with the DM and baryons well mixed.

We can define a projected mass-to-gas ratio, 
\beq
\upsh\equiv {{\Sigma} \over {m_{\rm HI} {\NHI}}} \,, 
\label{eq_upsh}
\eeq
which is determined by these different factors. 
Observations suggest values of $f_b\sim0.1$ \citep{kirk:00}, 
$f_{\rm H}\simeq0.75$, and ${\rm Y}_{\rm HI}$ in the range of 
$10^{-2}-10^{-3}$ for $z=0-5$ \citep{proc:99,hm:96}. 
Hence from Eqs. (\ref{eq_NHI}) and (\ref{eq_upsh}) 
we expect $\upsh$ to be in the range of $10^{2}-10^{4}$. 

In a similar manner it is possible to define a 
mass-to-ionic-column-density ratio that relates
the magnification to a metal line column density.
This introduces an additional parameter, the metallicity of the gas, 
or, more accurately, the specific abundance of a certain metal in a 
specific ionization state. Throughout this paper, except in 
\S \ref{sec_observe}, we will focus specifically on HI absorption. 
However, everything we will present can also be applied to any 
observed metal line.

With the above definitions we can now express the magnification of 
a quasar by an absorption system along its path as
\beq \label{eq_mag}
\mu=1+2{\upsh {\rm N}_{\rm HI} m_{\rm H} \over \Sigma_{cr}}.
\eeq
If we could measure the magnification of a single quasar,
we could determine $\upsh$ for every observed absorption system.  
However since there is no way to know a quasar's un-lensed luminosity,
we must exploit the correlation between magnification and the 
observable $\NHI$ to determine $\upsh$ statistically.
\footnote{Strictly speaking we determine the cross-section weighted 
average of $\upsh$.}

For a statistical sample of quasars, magnification has two effects.
It causes the observed solid angle to correspond to a smaller
solid angle in the source plane, and therefore the density of 
quasars with fluxes intrinsically above a flux limit decreases. 
However, magnification also increases the flux received from each 
quasar and therefore more quasars are detected above the flux limit.  
The net effect depends on the slope of the luminosity function of 
quasars. An increase in the number of quasars is detected if the slope
is steeper than $-2$ and a decrease otherwise. In the situation we 
will be considering, magnification will tend to increase the number of 
quasars brought into a sample.

In the next section, we illustrate the essence of the effect studied
here with a toy model. Most of its simplifications will be dropped in
later sections, to which readers familiar with gravitational lensing
might want to skip right away.

\section{Simple Model} \label{sec_sm}
In this section we introduce
a simple model that demonstrates the basic concepts of the method for
a statistical recovery of $\upsh$ without the complications that will 
be introduced in the following, more realistic case. 

Let us assume all quasars are at $z=3$ and their luminosity function 
(LF) is given by a power law of slope $-3.0$.  The number of 
absorption systems as a function of column density and redshift with 
column densities in the range 
$18 <\log\NHI<19.7$\footnote{Throughout this paper logarithmic
column densities are assumed to be measured in $\cm2$.}
is taken to be
\beq 
\label{eq_abs}
\rmd n=0.096 {\rm N}_{18}^{-1.5}(1+z)^{1.55}
\rmd {\rm N}_{18} \rmd z
\eeq
where ${\rm N}_{18}=\NHI/(10^{18} \cm2)$.
The choice of these parameters will be 
justified in section \ref{subsec_mc}. Throughout this paper we adopt 
the cosmological parameters
${\rm H}_0=65\, {\rm {km\,s}}^{-1} {\rm {Mpc}}^{-1}$, 
$\Omega_M=0.3$ and $\Omega_{\Lambda}=0.7$.  

Using these distribution functions, we construct a sample of quasars 
and the associated absorption line systems along their lines of sight.
The magnification value for each quasar is calculated according to the 
systems in its foreground, and the apparent luminosity of the quasar 
is correspondingly corrected by applying Eq. \ref{eq_mag}. 
The dilution of quasars by solid-angle magnification is taken into
account by randomly removing quasars. This is performed for values of 
$\upsh$ of $0$, $1000$, and $2000$ (we loosely denote the 
un-lensed case by $\upsh=0$). 

A flux cutoff at 45 $\mu$Jy (B brighter than $20^m$) 
is then applied to the magnified sample. Quasars which initially were 
below this flux limit are magnified by their LLS, exceed the flux 
limit, and introduce more high column density systems into the sample. 
\begin{figure}[h] 
\centering
\vspace{15pt}
\epsfig{file=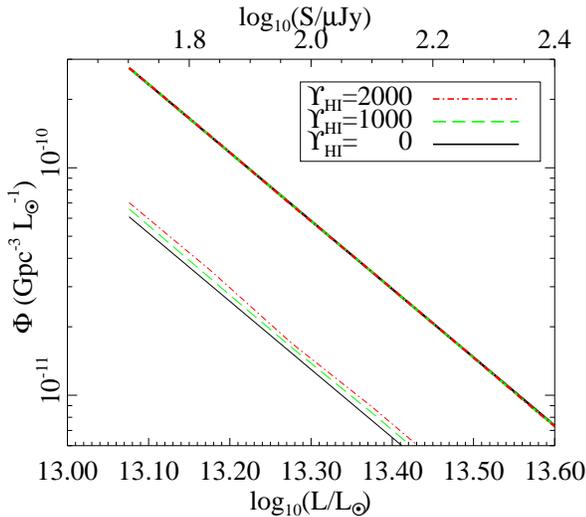,width=\linewidth}
\vspace{10pt}
\caption{The observed luminosity function of our toy model for all 
quasars and for quasars with at least one absorption system with 
$\log{\NHI} > 19.0$.  Each is plotted for three values of $\upsh$ 
(0, 1000, 2000).  The upper three lines lie on top of each other as 
weak lensing has a negligible effect on the total quasar luminosity 
function, as was previously found.  However the magnification
by gravitational lensing is discernible in the subset of quasars 
that contain high column density absorption systems.
}\label{fig_lf}
\end{figure}

Figure \ref{fig_lf} shows the quasar LF in this toy model.
The upper lines depict the LF of all quasars in the sample with
the three $\upsh$ values. Since the fraction of significantly magnified
quasars is negligible, the magnification (different $\upsh$ values) 
has no detectable effect on the total LF, in accord with the 
conclusions of \citet{pei:95b}.
The lower three lines depict the LF as derived from a subset of quasars
with at least one absorption system with $\log\NHI>19.0$.
For this subset it is evident that the LF is shifted to the right with 
increasing $\upsh$, i.e. increasing magnification.

In principle, the relative heights of the different lines in
Fig. \ref{fig_lf} are proportional to $\upsh$ and could be used to
find its value.  Unfortunately, in practice there is only one 
realization of the Universe so only one of the lines (for each $\NHI$ 
cutoff) in Fig. \ref{fig_lf} is observed.  In order to find out what
is the relative amplitude, the lowest line (un-lensed, $\upsh=0$)
should be inferred from the expected fraction of high column density
systems, based on lower column density systems and Eq. (\ref{eq_abs}).
Thus the effect of magnification is to change the fraction
of observed quasars with high column density absorption systems.   

\begin{figure} [h] 
\centering
\epsfig{file=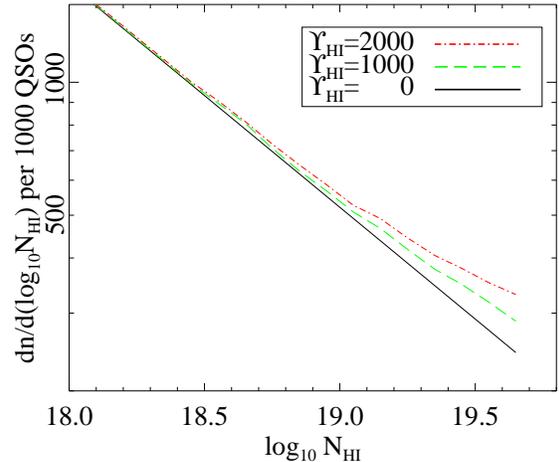,width=\linewidth}
\vspace{10pt}
\caption{The distribution of column densities of 
observed absorption systems for values of $\upsh$ of 0, 1000, and 2000.
The number of absorption systems observed per 1000 quasars is plotted 
versus the log of the column density.  For higher values of $\upsh$ 
more high column density systems are brought into the sample because 
of the magnification bias on the background sources.}\label{fig_N}
\end{figure}
This effect is more clearly seen if instead of looking at
the luminosity function we look at the distribution in column 
density of the absorption systems. Figure \ref{fig_N} shows this 
distribution for different values of $\upsh$.  The effect of 
increasing $\upsh$ is immediately obvious; more high column density 
systems are brought above the flux limit due to magnification bias. 
The distribution also deviates from the original power law as the 
effect becomes progressively more pronounced with increasing $\NHI$.  

\begin{figure} [t] 
\centering
\centerline{\epsfig{file=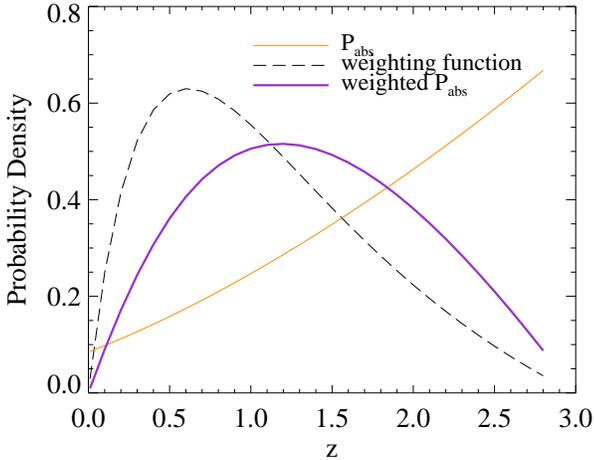,width=\linewidth}}
\vspace{10pt}
\caption{Magnification weighted probability of an absorption system 
as a function of absorption redshift (thick solid line) for quasars at 
$z=3$. The unweighted probability of an absorption system 
(Eq. \ref{eq_abs}, thin dotted line is weighted by $D_{\rm eff}$ of 
Eq. (\ref{eq_sc}) (dashed line).  All three functions are normalized
so that they integrate to unity.  The combination of magnification 
weighting which peaks at intermediate redshifts, and a monotonically
increasing probability of encountering an absorption system with 
redshift, results in a rather broad redshift range in which lensing by 
absorption systems is important. 
}\label{fig_pz}
\end{figure}

The magnification which causes the line divergence in Fig. \ref{fig_N}
is not only a function of $\upsh$ but also of the lens' redshift
through $D_{\rm eff}$ in Eq. (\ref{eq_sc}). 
We demonstrate this dependence in Figure \ref{fig_pz} where 
the probability of encountering an
absorption system (here normalized to integrate to unity)
as given by Eq. (\ref{eq_abs}) is weighted by the redshift
dependent factors of Eq. (\ref{eq_sc}) that contribute to the
magnification.  The resultant function is a modification of 
Eq. (\ref{eq_abs}) (again normalized to integrate to unity) with a 
broad distribution in redshift.
This means that lensing by absorption systems is relevant over a wide 
range of redshifts.

\begin{figure} [t] 
\epsfig{file=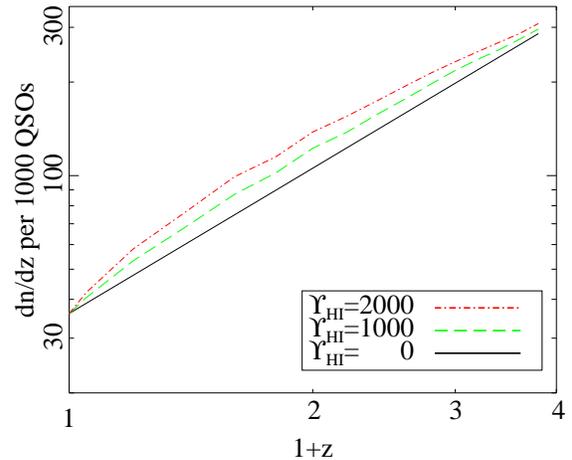,width=\linewidth}
\vspace{10pt}
\caption{The distribution of observed absorptions systems 
with column densities greater then $10^{19}\cm2$ in 
redshift for $\upsh$ values of 0, 1000 and 2000 for quasars at $z=3$.  
Since lensing depends on the redshift of the lens, systems that are 
at redshifts between 0.5 and 2.0 are preferentially brought into the 
sample. 
}\label{fig_z}
\end{figure}

This redshift dependence distinctly affects the counts of absorption 
systems per unit redshift for a fixed number of quasars. This effect 
depends on the value of the magnification coefficient, $\upsh$.
Figure \ref{fig_z} shows the number of absorption systems with 
$\log\NHI > 19.0$ as a function of absorption redshift.  
The increase in number of high column density absorption systems 
happens preferentially between redshifts of 0.5 and 2.0.  This is a 
clear sign that the increase is due to lensing for which the 
magnification becomes very small as the lens approaches the quasar 
redshift. A deviation from the redshift power-law dependence of 
Eq. (\ref{eq_abs}) is thus a statistical tool by which the amplitude 
of the magnification can be inferred.  These effects do depend to 
some extent on cosmology.

The toy model we have used here highlights the main effects of treating
the magnification by intervening absorption line systems in a 
statistical manner.  The luminosity function of quasars is shifted to 
higher luminosities, a shift that can be observed if a cut by column 
density along their line of sight is performed. However, it is 
cumbersome to get a reference line for a non-lensed luminosity 
function, whereas it is more straightforward to detect the change in 
the column density function and its redshift distribution.

\section{Recovery method for $\upsh$}
\label{sec_method}
The results of the toy model as presented in the last section portrayed
the basic effects of lensing on the statistical distributions of 
quasars and their absorption systems.
In reality, however, there are various deviations from the idyllic
picture of the toy model. In order to assess our ability to use these 
altered distributions for the recovery of $\upsh$ we need to take all 
of these complications into account. Only a detailed imitation of a 
realistic situation will allow us to draw conclusions 
regarding the success or failure of the recovery method.

The main complications that should be included in a more realistic 
picture are; 
\begin{itemize}
\item
The observed luminosity function of quasars resembles a double power law, 
rather than a single one, and thus an additional scale (the ``knee") 
is added to the problem \citep{bsp:88,hs:90,who:94,boyle:01}.
\item
The observed quasars span a large range of redshifts. This together with 
the redshift dependence of the LF ``knee'' results in a complicated 
{\it observed} flux function.
\item
An observed sample is flux limited, but quasars may be magnified above 
the flux limit depending on their absorption systems and the value of 
$\upsh$.
\item
Frequently, more than one absorption system intersects the line of 
sight to a quasar. These events, especially for high column densities, 
contribute a substantial signal to the shifts of the distribution 
functions. Multiple magnification events must thus be included in the 
analysis.
\item
Observational errors in $\NHI$ and flux measurements introduce various
biases to the method. These biases must be treated to evaluate their 
systematic effects on the recovery result.
\end{itemize}

For all these reasons one must appeal to Monte-Carlo realizations
of a realistic mock catalog of quasars and absorption systems.
The recovery method is demonstrated and tested using these 
Monte-Carlo realizations.

\subsection{The Mock Catalog} \label{subsec_mc}

We generate a sample of quasars according to their observed flux and
redshift distribution ${{{\rm d}n_{\rm QSO}}/{{\rm d}S{\rm d}z}}$.
Although this is the quantity which is directly observed, 
it is not reported in the literature. Instead it is converted to a 
luminosity function by binning under an assumed cosmology. We thus 
start with a fit to the luminosity function and recover the observed 
flux function. We start with the double power-law fit to the 
luminosity function as given by \citet{pei:95}.
\beq
\label{eq_phi_qso}
\Phi(L,z)={{{\rm d}n_{\rm QSO}}\over{{\rm d}L{\rm d}V}}=
{{\Phi_*/L_z}\over{(L/L_z)^{\beta_1}+(L/L_z)^{\beta_2}}}
\eeq
where $L_z=L_*\exp{[-(z-z_*)^2/(2\sigma^2)]}$ describes the redshift
evolution of the luminosity function, 
$\Phi_*=234\, {\rm (comoving) Gpc}^{-3}$, 
$\beta_1=1.83$, $\beta_2=3.70$, $L_*=2.63 \times 10^{13} L_{\sun}$,
$z_*=2.77$ and $\sigma=0.91$. 
We convert this to the observed flux function by
\beq \label{eq_lf}
{{{\rm d}n_{\rm QSO}}\over{{\rm d}S{\rm d}z}}=
4\pi{{{\rm d}n_{\rm QSO}}\over{{\rm d}L{\rm d}V}}
{{{\rm d}L}\over{{\rm d}S}}{{{\rm d}V}\over{{\rm d}z}}.
\eeq

For the probability of encountering an absorption system $P(\NHI,z)$
we assume this function is separable in $\NHI$ and $z$ as observations
seem to imply.
\citet{stor:94} found that the number density of LLS
increases as $(1+z)^{1.55}$.  Many authors 
\citep{tytler:87,pr:93,hu:95,lu:96b,kt:97,kim:97} showed that 
the distribution in column density is well fit by $\NHI^{-1.5}$ 
over ten decades. Thus we take the probability density to be
\beq \label{eq_pnz}
\rmd P= 
0.096 {\rm N}_{18}^{-1.5} (1+z)^{1.55} \, \rmd{\rm N}_{18} \, \rmd z
\eeq
with ${\rm N}_{18}=\NHI/(10^{18} \cm2)$ as in section \ref{sec_sm}.  
We only consider absorption systems with column densities between 
$10^{18} \cm2$ and $5 \times 10^{19} \cm2$. At the lower limit the 
magnification in the most favorable configuration (maximal 
$D_{\rm eff}$ and high
$\upsh$) is only $0.01$ so we can safely neglect lower column 
density systems when considering lensing.  

The upper limit is set because we would like all the systems we 
consider to contain gas in a similar ionization state 
(${\rm Y_{HI}}$), i.e, similar $\upsh$.
This upper limit must be set lower than the column density of 
damped \lya ($\NHI>2\times10^{20}\cm2$)
that consist of mostly neutral gas \citep{pw:96}.
A kinematic analysis of damped \lya by \citet{mpsp:01} 
suggests that cold gas disks at $z\simeq3$ extend down to 
$\NHI \sim 4 \times 10^{19} \cm2$, and local observations of the 
average surface densities of gas disks  yield
$\NHI \ga (few)\times10^{19}$ \citep{zwaan:97}.  
Taking inclination effects into account this implies 
that absorption systems with column densities below 
$5 \times 10^{19} \cm2$ are dominated by hot gas.
We therefore take this value as a conservative upper limit.  
This choice of an upper limit also safely keeps us 
away from strong lensing events that would not be described 
correctly by Eq. (\ref{eq_mag}).

Our results for $\upsh$ turn out to be quite insensitive to a
change in the assumed cosmology, the quasar flux function, 
or the less constrained column density function. 
Their primary effect is a slight modification in the number 
of quasars required to reach a result with the same level of confidence.
In light of the current uncertainty of over an order of magnitude 
in the value of $\upsh$, we will ignore these possible modifications.  

Armed with these relations we turn to generate a mock catalog of 
quasars and their associated absorption systems.  For the rare high 
(though still weak) magnification events which lift quasars above the 
flux limit of the survey, there is a need for a large sample of 
quasars.  We have generated a sample of $8\times10^6$ quasars and 
assigned  them fluxes and redshifts according to the flux/redshift 
distribution, ${{{\rm d}n_{\rm QSO}}/{{\rm d}S{\rm d}z}}$ 
(Eqs. \ref{eq_phi_qso}, \ref{eq_lf}).

In generating this catalog we have in mind the specifications of the 
SDSS and apply a flux limit of $45\mu$Jy to approximate 
the SDSS limit of $19^m.7$ in the $g'$ band. The target for the
photometric errors of the Sloan survey are $2\%$.  Based on what is 
commonly quoted in the literature we estimate that the errors in 
measuring $\NHI$ to be around $15\%$.
This catalog is used in the next section to exercise and test our 
proposed method for the $\upsh$ recovery.

\subsection{The Recovery Procedure} \label{subsec_p}

Our goal is to present an unbiased recovery method for $\upsh$,
assumed at first to be a universal constant. Relaxation of this 
assumption will be discussed in \S\ref{subsec_non_c_u}.
The recovery is judged by the confidence levels that can be put on
the value of $\upsh$ as function of its value and the number of
quasars above a flux limit that are needed in order to get these levels.
A magnification signal is manifested in the change of the distribution
functions of the column density: the relative number density of 
different column density systems and their redshift distribution. 
For $P(\NHI, z)$ there are three independent variables: 
the amplitude of the function, the slope of the separable $\NHI$ part 
of the function, and the slope of the separable $z$ part of it. 
Note that the amplitude can only be used once. We assume that the true
underlying slopes can be read off the low column density systems that
produce only negligible magnification, and then extrapolated to higher
column densities. Justifications for this assumption and ways to check
its validity will be presented in 
\S\ref{sub_checks}.
\begin{figure}[t] 
\centering
\centerline{\epsfig{file=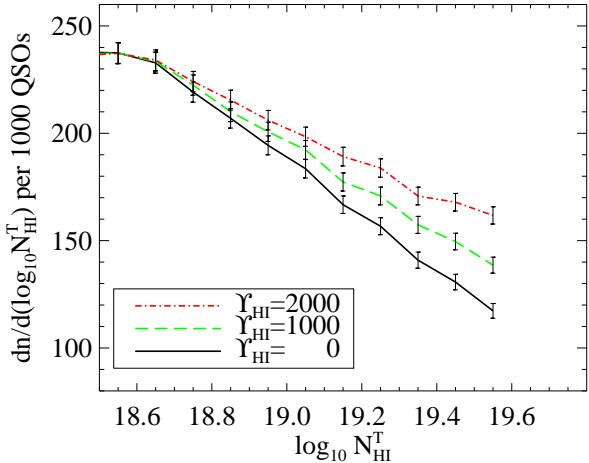,width=\linewidth}}
\vspace{10pt}
\caption{The distribution of the total projected HI
column density, $\nth$, for values of $\upsh$ of 0, 1000 and 2000.
Without lensing the distribution would be that of the $\upsh=0$
case.  Because of the magnification bias, systems with high values 
of $\nth$ are magnified over the flux limit. The error bars are 
the Poisson errors from a Monte-Carlo realization with $\nqso=10^5$.
}\label{fig_Nht}
\end{figure}

\subsubsection{The total column density along the line of sight}
\label{subsubsec_nth}

Let us now introduce a new variable, $\nth$, the total 
column density of HI along the line of sight. 
For the $j$-th quasar in the sample,
\beq
{\nth}_{j} = \sum_i {\NHI}_{j,i} \,,
\label{eq_def_nth}
\eeq
where the sum is over all systems along the line of sight.
Note that this is not weighted by $D_{\rm eff}$.
The probability distribution of this variable, $P(\nth)$, 
folds in the redshift/flux distribution of the quasars,
the column density/redshift distribution of the Lyman-limit systems, 
and in addition the Poisson process of multiple systems along the 
line of sight.

The total magnification experienced by a quasar is the product of the 
magnification by each absorption system along its line of sight.
Thus $\nth$ has a stronger correlation with 
magnification than the column densities of individual absorption 
systems. Another advantage of the $\nth$ quantity is that its 
measurement does not require assumptions on cosmology.
Figure \ref{fig_Nht} shows $P(\nth)$ for three values of $\upsh$.
There is a dramatic dependence on $\upsh$ in the distribution
$P(\nth)$ for higher total column densities.  
This should be compared to Fig. \ref{fig_N} where the effect of 
lensing was much less evident.
 
\subsubsection{Rejection of the null hypothesis}
\label{subsubsec_reject}

We first examine the rejection levels of the null hypothesis that the 
sample under consideration is not magnified at all. 
Under this assumption any deviation from the expected $P(\NHI, z)$ is
purely because of Poisson errors due to a finite sample and 
observational errors. As discussed above the relevant
distribution function to maximize the lensing detection is $P(\nth)$
rather than $P(\NHI, z)$. The equivalent projection along the $z$ axis,
$P_{\NHI}(z)=\int{\rm d}\NHI P(\NHI, z)$ can in principle also be used,
as we demonstrated in Fig. \ref{fig_N}.
Let us first phrase the statistics we use in the conventional $\chi^2$
terms and then modify it to get a statistic which is more
appropriate for the problem under consideration.

For a sample of $n_{\rm QSO}$ quasars, let $n^{\rm model}_j(\nth)$ 
be the number of expected absorber in bin $j$ spanning the range
$[\nth,\nth+\Delta\nth]$. This number depends, of course, on the
redshift distribution of the quasars in the sample.
Let $\tilde{n}^{\rm model}_i(z)$ be the number of expected absorbers 
in bin $i$ that spans the redshift range $[z,z+\Delta z]$.
If the two observed values of the same quantities are 
$n^{\rm obs}_j(\nth)$ and $\tilde{n}^{\rm obs}_i(z)$
we can construct the two $\chi^2$ functions
\begin{eqnarray}
\chi^2_1 &=& \sum_{j} { {\left[(n^{\rm model}_j(\nth) - n^{\rm obs}_j(\nth)\right]^2}
\over {\epsilon_{{\rm poisson},j}^2 + \epsilon_{{\rm error},j}^2}} \nonumber \\
\chi^2_2 &=& \sum_{i} { {\left[(n^{\rm model}_i(z) - n^{\rm obs}_i(z)\right]^2}
\over {\epsilon_{{\rm poisson},i}^2 + \epsilon_{{\rm error},j}^2}}\,,
\label{eq_chi2}
\end{eqnarray}
where the sums are over the relevant bins. 
The two error contributions are from the Poisson error due to the 
finite sample, $\epsilon_{\rm poisson}$ and the propagating error 
due to the observational uncertainty
in $N_{\rm HI}$. The slope in the $P(N_{\rm HI})$
function is affected by a Malmquist-like bias since it is more likely
for lower column density systems to ``leak" into a bin of higher column
density than the other way around.
We therefore adopted the following form for the error terms,
\begin{eqnarray} \label{eq_err}
\hspace{-.7cm}
&&\epsilon_{{\rm poisson},j} = \left(n^{\rm model}_j\right)^{-1/2} \\
\hspace{-.7cm}
&&\epsilon_{{\rm error},j} = 
{\frac{1}{2}{\delta{\rm N}_{\rm HI}}\over{\Delta\nth}}
\left(
n^{\rm model}_{j-1} - 2n^{\rm model}_{j} + n^{\rm model}_{j+1}
\right) \nonumber ,
\end{eqnarray}
where $\delta{\rm N}_{\rm HI}$ is the measurement error in $\NHI$ 
($15\%$).
We approximate the Poisson error from the modeled (expected) $n_j$
since ultimately we would like to get a probability distribution about
the true underlying value of $\upsh$.

The two functions, $\chi^2_{1,2}$ are not independent as written
in Eq. (\ref{eq_chi2}). If, for instance the total number of 
observed absorption systems exceeds the expected one, this will
affect both the $\chi^2$ values. In order to regain the statistical
independence, one of the $\chi^2$ functions should be renormalized such
that the modified function is read
\beq
{\tilde \chi}^2_2 = 
{\sum_{i}{\left({{n^{\rm model}_i(z)} \over{\sum_i n^{\rm model}_i(z)}}
-{{n^{\rm obs}_i(z)} \over{\sum_i n^{\rm obs}_i(z)}}\right)^2}
\over
{{\tilde \epsilon}_{{\rm poisson},i}^2 
+ {\tilde \epsilon}_{{\rm error},i}^2}} \,,
\eeq
where the original errors, $\epsilon$, are re-scaled accordingly to 
${\tilde \epsilon}$.  The final $\chi^2$ function is then expressed 
as the weighted sum of the two independent functions
\beq 
\label{eq_chi2_both}
\chi^2 = {c_1\chi^2_1 + c_2{\tilde \chi}^2_2}\,;\,\, c_1+c_2=1\,.
\eeq
In practice we find that the statistics is dominated by $\chi^2_1$ and 
little improvement in the statistical rejection levels or confidence
levels is obtained by including ${\tilde \chi}^2_2$. We therefore set
$c_1=1$ and $c_2=0$ for the rest of our analysis.
For example, the $\chi^2$ statistics allows us to reject the null 
hypothesis with $\nqso=2 \times 10^4$ and values of $\upsh$ of 300, 400
and 500 at the $70\%, 82\%$ and $90\%$ confidence levels respectively.

So far we have used standard $\chi^2$ statistics.  However, in the 
current case, where we try to detect a non-random process (or to 
reject the absence thereof) there is a better, more robust way of 
doing it. Based on Fig. \ref{fig_Nht} we notice that the high column 
density end of the $\nth$ distribution can be well fit by a power law.
This power-law description is excellent for the non-magnified case, but
it provides a reasonable description for the magnified cases as well.
A full description of the fit to the non-lensed case is provided 
by its normalization, $a_0$, and its slope $a_1$. 
Recall that since $\nth$ is the (plain, unweighted) sum of the 
$\NHI$ of all
absorption systems on the line of sight to a quasar, $a_1$ is {\it
not} identical to the slope of ${\rm d}P/{\rm d\NHI}$ of Eq. 
(\ref{eq_abs}). 
The function $P(\nth)$ is established using $10^6$ un-lensed quasars from
the mock catalog.  From this we determine $n^{\rm model}_j(\nth)$ 
for a given value of $\nqso$ and bin size.
The best fit to the values of $n^{\rm model}_j(\nth)$ 
and the errors on these values as given by equation
(\ref{eq_err}) provides the value of $(a_0,a_1)$.

In the presence of magnification, we expect a coherent deviation from
the non-magnified line ($a_0,a_1$) of the $\nth$ distribution.
This altered distribution can be fit  
with the two parameters which describe it best, ($a'_0,a'_1$), 
by a standard fitting procedure, which also
takes into account an error covariance matrix ${\bf \Delta a}$ which 
is a function of $n^{\rm model}_j(\NHI)$ and of the $\epsilon_j$s.
\begin{figure} [t] 
\epsfig{file=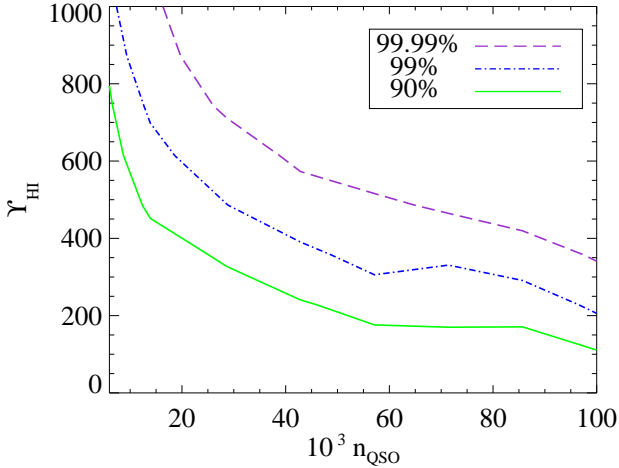,width=\linewidth}
\vspace{10pt}
\caption{Rejection levels for the null hypothesis
that the observed distribution of $\nth$ is unbiased by gravitational
lensing as a function of the number of quasars in the sample and the
value of $\upsh$. $\nqso$ is the total number of quasar in the sample,
but only a subset of these quasars are used in order to calculate the
rejection levels. The subset is determined by a cut in
$\nth$ so that $19.0<\log\nth<19.7$.
This amounts to $10\%$ of
the $n_{\rm QSO}$ in the case of no lensing and increases to $12\%$ 
if $\upsh$ is 1000.}
\label{fig_r19}
\end{figure}

At this point the new statistic we establish compares two lines; the
model line ($a_0,a_1$) and the fit to the observed line ($a'_0, a'_1$).
We use  Monte-Carlo runs to map the probability space in the 
parameters ($a_0, a_1$) of the non-magnified $\nth$ distribution and 
to evaluate the probability of fitting a line (in log) of 
normalization $a'_0$ and slope $a'_1$ given a random, discrete 
realization of this distribution with $\nqso$ lines of sight.
The outcome is a full map of the distribution
${\rm d}P/({\rm d}a_0 {\rm d}a_1)$ which is similar to
Eqs. (\ref{eq_chi2}) replacing $n^{\rm model}_j(\NHI)$ by the fit
to the modeled $\nth$ and $n^{\rm obs}_j(\NHI)$ by the fit to the
realization of the $\nth$ distribution. Under this change of variables
there is no longer summation over bins and the function which
effectively replaces $\chi^2$ becomes,

\beq
\psi= {\vert (a_0,a_1) - (a'_0, a'_1) \vert ^2
\over \vert {\bf \Delta a} \vert} \,.
\eeq

For the z distribution this statistic cannot be used because 
magnification causes an increase of the number of systems in some 
bins and a decrease in others.  Thus a magnified distribution is no 
longer well fit by a line (in the log). To compare instead an 
observed $z$ distribution with a non-magnified one, we use the 
standard Kolmogorov-Smirnov test. 
\begin{figure} [t] 
\epsfig{file=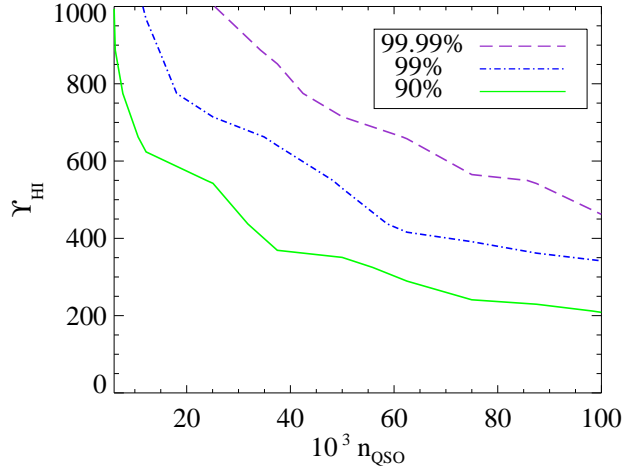,width=\linewidth}
\vspace{10pt}
\caption{Same as Fig. \ref{fig_r19}, but rejection levels are drawn 
based only on quasars with $19.5<\log\nth<19.7$ which corresponds to 
$2.1\%$ and $2.7\%$ of $\nqso$\, for $\upsh=0$ and $\upsh=1000$ 
respectively.
}\label{fig_r19.5}
\end{figure}

Based on this $\psi$ statistic we can draw rejection levels of the
null hypothesis. These are more stringent than those obtained
using $\chi^2$; for example with a sample of $2\times 10^4$ quasars an
$\upsh$ of 600 can be rejected with $99.4\%$ confidence using the
$\psi$ statistic, but with only $96\%$ confidence using $\chi^2$.  
Figure \ref{fig_r19} shows the rejection levels as function of $\nqso$
and the true modeled value of the constant $\upsh$. Even though 
$\nqso$ is of the order $10^4-10^5$ ($x$ axis), the rejection levels 
are drawn from a subset of quasars that were selected based on their 
$\nth$ value. This subset amounts to $10\%$ of the quasars if 
$19<\log\nth<19.7$ and $\upsh=0$ and increases to $12\%$ for 
$\upsh=1000$.
As expected, the rejection confidence increases with increasing $\nqso$
and $\upsh$. For example, it is enough to have a total sample of
$2\times 10^4$ quasars in order to reject the null hypothesis by more 
than $99\%$ for a true $\upsh$ value of $600$.
Recall that only about one tenth of this sample is actually being used
in this calculation.

The fraction of the sample that is actually being used can be further 
reduced by a factor of $4-5$ if a higher lower limit for $\nth$ 
is adopted. Figure \ref{fig_r19.5} shows the rejection levels if 
only quasars with $19.5<\log\nth<19.7$ are taken into account. This 
reduces the sample to $\sim2.1\%$ and $\sim2.7\%$ of the size of 
the original sample for $\upsh=0$ and $\upsh=1000$ respectively.
However it now takes about $4.5\times10^4$ quasars in the original 
sample in order to reject the null hypothesis at the $99\%$ 
confidence level for a true $\upsh$ value of $600$.

\subsubsection{Confidence levels for recovered $\upsh$}
\label{subsubsec_confidence}

Rejection of the null hypothesis is only the first step in establishing
the value of $\upsh$ and its functional shape.
Similarly to the choice of $c_2=0$ in the $\chi^2$ statistics of the
previous section (\S\ref{subsubsec_reject}, Eq. \ref{eq_chi2_both}),
we only consider the change in the number counts in $\nth$ bins since 
the information in the change of the redshift distribution does not 
improve the confidence levels by more than a few percent
\footnote{In addition to the change in the absorbers' redshift
distribution, there is also a small shift in the redshift distribution
of magnified (high ${\nth}_j$) quasars relative to the non-magnified 
quasars.}.

To get the best value of $\upsh$ we de-magnify
each quasar in the sample individually, based on the {\it actual}
absorption systems along its line of sight. The observed flux, 
$S'_j$ of a quasar $j$ at $z_j$ is shifted to the de-magnified flux, 
$S_j$ by
\beq
S'_j \rightarrow S_j = {S'_j \over \prod_{i} \mu_{j,i}(z_j,z_i,\NHI^i;\upsh)}
\,,
\eeq
where $\mu_{j,i}$ is the magnification due to the absorption system $i$
of column density $\NHI^i$ at $z_i$ on the line of sight to the 
quasar, according to Eq. (\ref{eq_mag}).  If $S_j$ turns out to be 
lower than the ``survey" flux limit, then quasar
$j$ along with all its associated absorption systems are removed from
the sample. The remaining sample contains \npqso $(\leqslant \nqso$) 
quasars with a set of \npqso values of $\nth$.  Then in order to 
account for the dilution of quasars due to solid-angle magnification, 
we add quasars in the following way. Quasars are drawn randomly from
the remaining sample with a probability $1/\mu$ and then added to the 
sample such that there are duplicates of some of the quasars and their
$\nth$ values. These values are fit (in the log) to a linear function 
characterized by $(a'_0,a'_1)$ and compared to the non-magnified 
(i.e., the extrapolation from the low $\nth$ range) distribution fit.
This is done in the same manner described in \S\ref{subsubsec_reject}. 
The best $\upsh$ value is obtained by minimization of the $\psi$ 
statistic. 

\begin{figure} [t] 
\hskip-0.3truecm\centerline
{\psfig{file=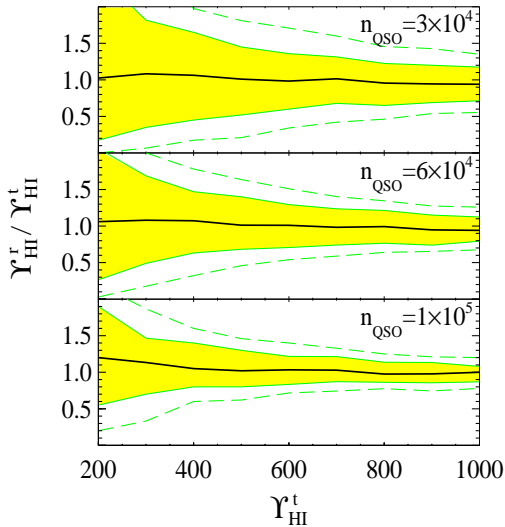, height=7truecm,width=9truecm }}
\vspace{10pt}
\caption{The ratio of the recovered ($\upsh^r$) to the true 
($\upsh^t$) value of the mass-to-gas ratio and the $95\%$  
(shaded region) and $99.7\%$ (dashed line) confidence regions are 
shown versus $\upsh^t$ for three quasar sample sizes. 
}\label{fig_recover}
\end{figure}

Figure \ref{fig_recover} shows the success of the recovery of $\upsh$
values along with its confidence levels.  The ratio of the recovered 
value  ($\upsh^{r}$) and the true pre-assigned value ($\upsh^t$) 
is shown as a function of $\upsh^t$.
Only quasars with $19.0<\log\nth<19.7$ were used to draw the confidence
levels.  As the total $\nqso$ increases (from the top panel down) 
the confidence levels of the recovery for a given true $\upsh^t$ value 
become narrower about this value. Note that the average recovered value
is not biased even when the confidence levels are broad.

All quantities are calculated using between $10-20$ Monte-Carlo 
runs depending on the value of $\nqso$. With a sample of $10^5$ 
quasars it is possible to measure $\upsh$ to a higher accuracy than 
$30\%$ of its value if it is greater than 500.
This is a remarkable improvement over the current uncertainty 
of an order of magnitude.
The uncertainty in the recovered value of the mass-to-gas ratio
does not depend on $\upsh^t$ but only on $\nqso$ (the relative 
uncertainty as shown in Fig. \ref{fig_recover} goes down because 
$\upsh^t$ increases). The $95\%$ uncertainty in the recovered value 
of $\upsh$ is approximately equal to 
$140 \, \sqrt{{10^5}\over{n_{\rm QSO}}}$.

Using only quasars with $19.5<\log\nth<19.7$ out of a sample size of 
$10^5$ quasars has the effect of broadening the confidence
levels to the degree of confidence level widths close to the case of
$\nqso=6\times10^4$ with $19.0<\log\nth<19.7$. 

\subsection{Non-constant $\upsh$}
\label{subsec_non_c_u}

\begin{figure}[t] 
\hskip-0.3truecm\centerline
{\psfig{file=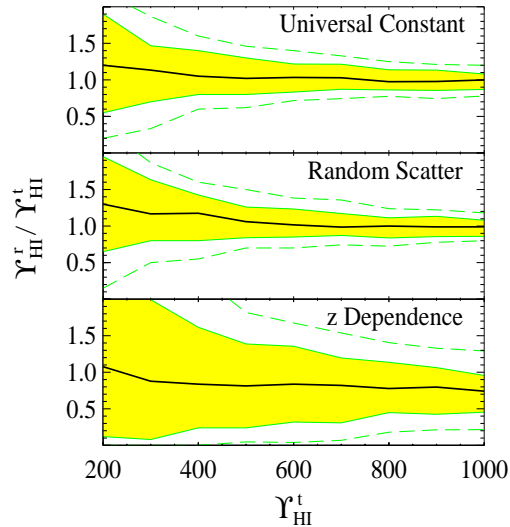, height=7truecm,width=9truecm }}
\vspace{10pt}
\caption{The ratio of the recovered ($\upsh^r$) to the true
($\upsh^t$) value of the mass-to-gas ratio for different mass-to-gas
ratio schemes.  The top panel is the universal constant scheme explored
in \S \ref{subsec_p}.  The middle panel shows the result from 
a random mass-to-gas ratio with a mean of $\upsh^t$ and a 
$\sigma_{\upsh}=\frac{1}{2}\upsh^t$.  
The bottom panel shows the results from the 
redshift dependent scheme explained in the text.  For all three panels
$\nqso=10^5$ and quasars with $19.0<\log\nth<19.7$ are used with the 
$\psi$ statistic. The shaded region and dashed line denote $95\%$ and
$99.7\%$ confidence levels respectively.
}\label{fig_random}
\end{figure}

The assumption of a universal mass-to-gas ratio is quite simplistic. 
We here explore two deviations from this assumption.
The first scheme we investigate assumes that $\upsh$ has a Gaussian 
distribution with mean $\upsh$ and standard deviation, 
$\sigma_{\upsh}$. The middle panel of Figure \ref{fig_random}
shows the recovered value for $\upsh$ and the confidence levels about 
it for $\sigma_{\upsh}=0.5\upsh^t$.
Introduction of this significant scatter in the mass-to-gas ratio has
a modest effect on the recovery of $\upsh$.  The recovered mean value 
is slightly higher than $\upsh^t$, but this
small bias is less than $50\%$ of the uncertainty of the recovered
value of $\upsh$ even for values of $\sigma_{\upsh}$ as large as
$\upsh^t$.  Thus it is only noticeable for small values of $\upsh^t$ 
as can be seen in Fig. \ref{fig_random}.
Part of this bias arises because $\upsh$ must be positive, which 
skews the Gaussian distribution, thereby increasing
the mean value of $\upsh$. However, there is also the effect
that in an observed sample higher values of $\upsh$ will preferentially
be brought into the sample, increasing the mean $\upsh$ value of 
observed absorption systems. 
The scatter in the recovered value shows no noticeable increase for 
$\sigma_{\upsh}=0.5\upsh^t$ because the Poisson scatter in the 
distribution of absorption systems dominates the noise. 
Thus the statistical nature of the recovery and the large data set 
being used makes random scatter unimportant in the recovery of $\upsh$.

The second scheme for the deviation of $\upsh$ from a universal 
constant is motivated
by the ionization evolution of the Universe. If the ${\rm Y}_{\rm HI}$
parameter of LLS in Eq. (\ref{eq_NHI}) is a function of redshift
then so is $\upsh$.
For the functional form of ${\rm Y}_{\rm HI}$ we adopt \citep{hm:96}
\beq
{\rm Y}_{\rm HI} = { \rho_{\rm H} \alpha_{\rm HI}(T) \over \Gamma_{\rm HI} -
\rho_{\rm H}\alpha_{\rm HI}(T) } \,
\eeq
where $\Gamma_{\rm HI}$, the photo-ionization rate per hydrogen atom, 
is given by
\beq
A(1+z)^B\exp\left[ - {(z-z_c)^2 \over s } \right] \,,
\label{eq_Gamma}
\eeq
with the fitting parameters $A=7\times10^{13}$, $B=0.8$, $s=1.95$, and
$z_c=2.2$.  $\alpha_{\rm HI}$ is assumed to be $2.6\times 10^{-13}$, 
its value for a $10^4\,$K, optically thin gas, with density of 
$\rho_{\rm H}=4\times 10^{-3}$cm$^{-3}$. This roughly corresponds to 
the virial density and $f_b=0.1$. In practice we adjust $A$ so that 
the maximal $\upsh$ value we obtain is $3000$.

First, we address the issue of recovering $\upsh$ under the assumption
that it is a universal constant.  In this case we will refer to the 
true value of the mass-to-gas ratio $\upsh^t$ as the average value 
of $\upsh$. For the parameters given above the average value is 
$42\%$ the maximal value.

When $\upsh$ really does vary with redshift, the
bottom panel of Fig. \ref{fig_random} shows the results of the
recovery method.  The recovered value is about 2/3 the
average value which reflects the sensitivity of lensing to redshift 
as shown in Fig. \ref{fig_z}. Also the uncertainty in the 
recovered value is much wider then in either the universal constant or 
random scatter schemes.

\begin{figure} [t] 
\epsfig{file=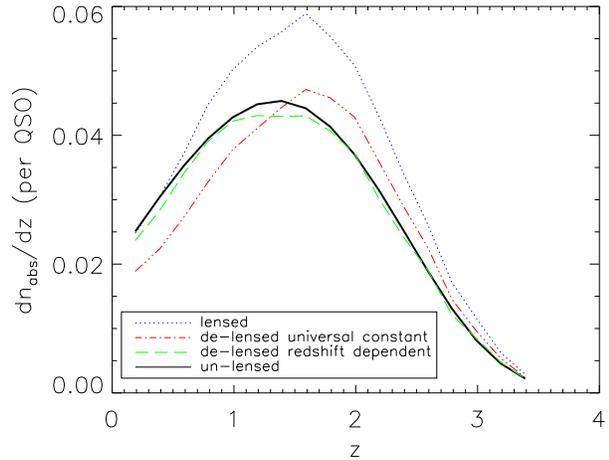,width=\linewidth}
\vspace{10pt}
\caption{The number of absorption systems with $\log{\NHI} > 19.0$ 
as a function of redshift.  The thick solid line is the
un-lensed case.  The dotted line is the lensed sample magnified by a 
mass-to-gas ratio that depends on redshift (see text). 
The dot-dot-dot-dash line is obtained when the sample is
erroneously de-magnified by an assumed universal constant
value for $\upsh$. 
De-magnifying by the correct $z$ dependent scheme results in the 
dashed line. 
}\label{fig_zdis}
\end{figure}

The second issue of concern is whether we can determine $z_c$ 
in Eq. (\ref{eq_Gamma}) using our recovery method.  First we need
to know that $\upsh$ is a function of $z$ and not a universal 
constant.  This can be done by examining the distribution in redshift
of absorption systems with $\log{\NHI} > 19.0$ in the demagnified 
sample.  Figure \ref{fig_zdis} shows that while the 
demagnified sample fits the unbiased distribution of $\NHI$ the $z$
distribution of its absorption systems is markedly different.

\begin{figure} [t] 
\epsfig{file=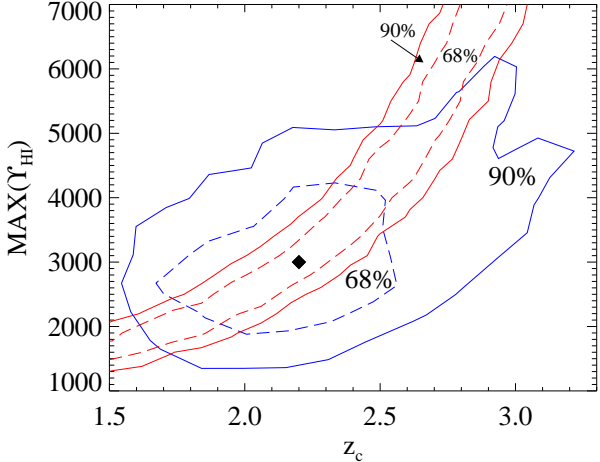,width=\linewidth}
\vspace{10pt}
\caption{The ability to recover the values of $z_c$ and the maximum
value of $\upsh$ using the $\nth$ distribution and the $z$ distribution
of absorption systems. The broad, closed region
shows the constraints from considering
the $z$ distribution and the narrow arc comes from considerations of
the $\nth$ distribution. This is done with $\nqso=10^5$. 
Both tests individually allow for a large range in $z_c$ but 
their combination can yield a fairly good measure of these parameters.
}\label{fig_zrecover}
\end{figure}

Finding that the assumption of a universal constant $\upsh$ provides 
a poor fit to the expected $z$ distribution of absorption systems with 
$\log{\NHI} > 19.0$, we can try other mass-to-gas schemes.  If we 
assume a function of the form Eq. (\ref{eq_Gamma}) we can allow $z_c$ 
and the maximum value of $\upsh$ to vary and compare the resulting
de-magnified samples against the expected $\nth$ and $z$ distributions.
The results are shown in Figure \ref{fig_zrecover} where the $\psi$
statistic has been used to compare to the $\nth$ while a K-S test is 
used to determine the probability that the demagnified $z$ distribution
is compatible with the expected one. The narrow curve is the 
constraint from the $\psi$ statistic while the broader, closed 
region shows the constraints
from matching the $z$ distribution.  The black diamond shows the true
values used in the magnification.  Values of $z_c$ that
are compatible with both tests at better then the $90\%$ level lie
between redshifts of 1.5 and 2.9.  
Thus constraints on the redshift dependence of $\upsh$ can be placed 
by considering the $z$ distribution of absorbers in combination with 
the $\nth$ distribution. 

\subsection{Additional checks for the method}
\label{sub_checks}

Checks that the signal measured and ascribed to
gravitational lensing is really due to gravitational lensing come
naturally in our scheme. The first verification
is to check that with the recovered value of $\upsh$ the demagnified
sample is in all ways unbiased.  The second is to break the observed 
sample into a part favorable to being lensed and an unfavorable part 
and then perform the above procedure comparing these two parts.

The correct value of $\upsh$ should not only bring the distribution of
$\nth$ into agreement with what is expected for an un-lensed 
population, but also the flux function of these demagnified quasars 
and the redshift distribution of their absorption systems should be in
agreement with those quasars that are not significantly lensed.  
The ability of one value of $\upsh$ to bring the quasar sample into 
agreement with these three independent distributions seems quite 
unlikely if the measured difference in $\nth$ was due to some other 
effect than lensing.  If the $z$ distribution does not match the 
un-lensed one this does not rule out lensing, it simply rules out 
the dependence of $\upsh$ on redshift that we have assumed.  
A different dependence that is compatible with the three distributions
must then be found as has been demonstrated above. 

A second check is to divide the sample into those 
absorber-quasar pairs that
are favorable and unfavorable to lensing.  Equation \ref{eq_sc} shows
that lenses of the same mass with redshifts near the redshift of the 
quasar produce less magnification.  Thus separating the sample by this
criterion creates a subsample that is in every way identical to 
the total sample except for this one relationship {\it {that should
only be relevant to gravitational lensing}}.  The $\nth$ distributions
of the two samples can be compared and $\upsh$ recovered in exactly 
the same method as explained above.  The advantage of this
test is that it can be made solely with the data at hand with
no need to extrapolate $P(\NHI,z)$ from lower column densities; 
however, because one divides the observed sample into two parts the 
signal will be reduced.  Thus our preference is to assume that we can 
extrapolate $P(\NHI,z)$, and then perform the separation of the sample
as an additional check that our assumption is correct.  
In practice one would first want to validate the assumption that
$P(\NHI,z)$ can be extrapolated to higher column densities by checking
that those systems unfavorable to lensing are consistent with the
lower column density $P(\NHI,z)$.

\section{Observational procedure, metal lines, and dust}
\label{sec_observe}

The basis of any future large quasar study will be based on the 
five band photometry of $10^6$ quasars and spectra of $10^5$ quasars 
obtained by the SDSS
\footnote{http://www.astro.princeton.edu/PBOOK/welcome.htm}.
The Sloan spectra will identify \lya, MgII and CIV lines in the
redshift ranges, $2.3<z<6.4$, $0.5<z<2.2$, and $1.6<z<4.8$, 
respectively.

The detection of \lya from the ground is restricted to $z>2$ so only 
about $15\%$ of the absorption systems we have been considering in \S
\ref{sec_method} will be detected in \lya in the Sloan data.  
Furthermore these systems will be less favorable to lensing because 
of the proximity in redshift of the quasar and the absorber.
However, as shown in Fig. \ref{fig_r19}, if the true mass-to-gas
ratio exceeds 800 then with only 5000 quasars it is possible to reject
$\upsh=0$ (no lensing) at the $90\%$ confidence level.
So it is possible that a lensing signal may be detected with just the 
LLS identified in the Sloan survey.  

MgII, CIV and other metal lines are detectable over a wide range of 
redshifts from the ground and thus from the Sloan data alone values 
for $\Upsilon_{\rm MgII}$ and $\Upsilon_{\rm CIV}$ can be determined.  
While the theoretical interpretation of these values is more uncertain
than $\upsh$ because of the added complication of metallicity of the 
gas, nonetheless metal lines can establish that there is lensing by 
absorption systems. The mass-to-metal-line ratios of different elements
can be used to study the enrichment history of LLS.  
Using correlations between metal line and \lya column density, 
$\upsh$ can also be inferred, though with a bigger scatter.

The optimal sample consists of LLS below redshift 2.  
The Cosmic Origins Spectrograph (COS)
\footnote{http://cos.colorado.edu/cos} to be installed on the Hubble 
Space telescope and the Galaxy Evolution Explorer (GALEX)
\footnote{http://www.srl.caltech.edu/galex} will both be able to detect
\lya in the ultraviolet starting in 2003.  
Of course these missions will not be able to get spectra for $10^5$ 
quasars, but as we have described in \S \ref{sec_method} we only 
use the $10-12\%$ of quasars with $\log \nth >19.0$ in measuring the 
lensing signal and recovering $\upsh$. Thus if the quasars likely to 
cross this threshold can be identified from their MgII and/or CIV 
features in a manner identical to that already done when looking for 
DLAS \citep{rt:00} it only requires $\approx 10^4$ spectra to measure 
$\NHI$ in the relevant systems.  Since GALEX intends to get spectra 
for $10^4$ quasars it is even feasible that almost the entire Sloan 
sample can be used in measuring the value of $\upsh$.
Because the method of targeting MgII absorbers is the most effective 
way to detect DLAS with $z<2$, it is likely that a significant
fraction of telescope time will be spent on such an endeavor.  
About $5-10\%$ of all quasars are expected to have at least one
absorption system that is above our upper bound on $\NHI$ 
($5 \times 10^{19} \cm2$) and therefore would not be included in the
lensing analysis.  Of MgII absorbers $ \approx 50\%$ are above this
upper bound but the rest  will be exactly the higher
column density LLS that are needed for the lensing analysis.  Thus the
prospects of having UV spectra for a large number of quasars with 
$19.0<\log\nth<19.7$ is reasonably good.
 
One possible impediment to the method we have described in this paper
would be the presence of dust in LLS.  Metallicity is usually 
believed to be connected with the dust content in galactic systems. 
If in spite of their high ionization states, the LLS do contain dust,
then dust absorption may be proportional to $\NHI$ and may act as a 
counter effect to the gravitational magnification. A straightforward 
way to probe whether absorption by dust is important would be to apply
the recovery method from photometry at different band filters. Similar
values for $\upsh$ from analyses in different bands should verify the 
irrelevance of dust reddening.  If, however, a filter dependent result
is obtained for $\upsh$ values, this might be an extension of the 
method which would constrain the dust content in LLS.

\section{Conclusions} \label{sec_conc}

We have introduced a new statistical method to recover the mass to
neutral hydrogen ratio in Lyman limit systems. The method makes use of
the connection between the measured column density of individual
absorption systems and their contribution to the magnification of their
background quasars.

The main effect of the gravitational magnification is the change of
slope at the high column density end of the column density distribution
function. This can be also seen as a shift in the 
quasar luminosity function, a shift which depends on column densities. 
Accompanied effects include a distortion of the redshift distribution 
for the LLS and a small shift in the quasars redshift
distribution for quasars with high column density of total absorption.

The relevant variable for these effects is the total column density 
in LLS along the line of sight to a quasar. We have simulated in detail
the quasar population along with their associated LLS. 
We have mimicked the observational procedure and demonstrated the 
ability to recover the right value of the mass to neutral gas ratio, 
$\upsh$.  If $\upsh$ is a universal constant in these systems then the
$95\%$ confidence level of the recovered $\upsh$ value is given by
$\pm140 \left(\sqrt{{10^5}\over{{n_{\rm QSO}}}}\right)$ 
for a sample size in the range $10^4<n_{\rm QSO}<10^5$, 
independent of the value of $\upsh^t$! For a stochastic $\upsh$
with a Gaussian distribution with standard deviation $\sigma_{\upsh}$,
the $95\%$ confidence levels increase only slightly because the large
number of quasars diminishes the effects of stochastic processes.
A small bias is also introduced because observed higher values of 
$\upsh$ are preferentially brought into the sample and because the 
mean of an unbiased sample is also slightly higher since $\upsh$ is 
always positive.
However this bias is relatively small; for the large scatter of
$\sigma_{\upsh} = \upsh$ the bias is only half as large as the
$95\%$ confidence level.

A redshift dependence of the $\upsh$ value due, e.g., to evolution 
in the ionization state of the LLS can be recovered to a certain 
extent by assuming the functional form of the dependence and deriving 
the free parameters of that form.  For a reasonable example of a 
redshift dependent model \citep{hm:96}, the maximum ionization 
redshift can be recovered with an accuracy of $\pm 0.7$
These scaling relations for the recovery ability are all achievable in
light of the upcoming quasars survey of the SDSS. Follow up 
observations from space should then be undertaken in order to fill 
in the low redshift range of LLS, unaccessible from ground.

So far the mass to neutral gas ratio in LLS is unknown to more than an
order of magnitude. The method we presented here enables, for the first
time, to constrain this value tightly,
hopefully leading to a better understanding of the relation between
gas and dark matter in the universe.

\acknowledgments We would like to thank Tal Alexander, 
Joel Primack and Jason Prochaska for useful conversations. 
This work was supported in part by GIF, the NSF, NASA, 
the US-Israel Binational Science Foundation and a UCSC Faculty 
Research grant. 

\bibliographystyle{apj}        

\bibliography{lylens}

\end{document}